\documentclass[english,aps, pra, amsmath,amssymb, reprint]{revtex4-1}
\usepackage[T1]{fontenc}
\usepackage[utf8]{inputenc}
\usepackage{amsmath}
\usepackage{amssymb}
\usepackage{graphicx}
\usepackage{esint}

\makeatletter
\usepackage{slashed}

\makeatother

\usepackage{babel}
\begin{document}

\title{Analytically continued physical states in the path-integral: a sign-problem-free
Quantum Monte Carlo simulation of Bell states dynamics.}

\author{Evgeny A. Polyakov}
\email{e.a.polyakov@gmail.com}
\affiliation{Russian Quantum Center, Novaya 100, 143025 Skolkovo, Moscow Region, Russia}
\affiliation{Faculty of Physics, Saint Petersburg State University, 7/9 Universitetskaya Naberezhnaya, Saint Petersburg 199034, Russia}

\author{Alexey N. Rubtsov}

\affiliation{Russian Quantum Center, Novaya 100, 143025 Skolkovo, Moscow Region, Russia}
\affiliation{Department of Physics, Lomonosov Moscow State University, Leninskie gory 1, 119991 Moscow, Russia}

\begin{abstract}
The derivation of path integrals is reconsidered. It is shown that
the expression for the discretized action is not unique, and the path
integration domain can be deformed so that at least Gaussian path
integrals become probabillistic. This leads to a practical algorithm
of sign-problem-free Monte Carlo sampling from the Gaussian path integrals.
Moreover, the dynamical influence of Gaussian quantum system (the
bath) on any other quantum system can be exactly represented as interaction
with classical non-Markovian noise. We discuss the relation of these
findings to the Bell's theorem and the Feynman's conjecture on the
exponential complexity of the classical simulation of quantum systems.
In Feynman's path integral we have quasiprobability distributions
for trajectories, and in analitycally continued path integrals we
have probability distributions for quasitrajectories.
\end{abstract}

\maketitle

\section{INTRODUCTION}

One of the most characteristic features of quantum theory is that
it cannot be interpreted as a local causal (possibly stochastic) real-time
evolution in a certain space of ``hidden parameters'' \citep{Feynman1981}.
This is the content of Bell's theorem: we cannot have simultaneously
locality, causality, physical probabilities, and physical states.
Something should be sacrificed. Feynman's path-integral interpretation
of quantum mechanics succeeds in representing evolution as statistics
of diffusive trajectories (field configurations) $x$. To satisfy
the Bell's theorem, this interpretation sacrifices the physicality
of probabilities: they are no longer real positive but arbitrary complex
weights $\exp\left(-iS\left[x\right]/\hbar\right)$, where $S\left[x\right]$
is the action for configuration $x$. This is not so important for
analytic calculations, but becomes crucial for the numerics.

The physicality of trajectory probabilities is crucially important
in quantum Monte Carlo (QMC) simulations. The QMC is the only choice
in cases when it is difficult to identify small or large parameter,
and the dimension of phase space is high. A number of efficient QMC
algorithms are known for the imaginary time domain, so that contemporary
QMC can efficiently solve equilibruim quantum problems. However in
real time the trajectory weights are complex and oscillating. Currently
there is no understanding of how to efficiently arrange the sampling
procedure. This is the so-called sign problem. 

For more than half a century, it has been made a lot of attemps to
solve this problem. The most notable and oldest are: to enforce the
controlled cancellation of oscillating trajectory weigths (filtering
\citep{Filinov1986,Makri1987,Makri1988} and multilevel blocking \citep{Mak1999,Mak2009}
techniques); analytic continuation with respect to the time variable
$t$ \citep{Creutz1981,Bonca1996,Ramirez2001}. Nevertheless, none
of these methods enables us to simulate large correlated systems for
long enough times. 

Another group of approaches which is currently under active investigations
is based on the idea to perform analytic continuation of probability
amplitudes with respect to the (field) configurations $x$. This approach
takes its origin from the works of Parisi \citep{Parisi1983} and
of Klauder and Petersen \citep{Klauder1985} where first time the
idea was proposed of how to sample the complex probabilities with
the aid of complex Langevin equation. While having limited success
for certain model systems, the method of complex Langevin equations
suffer from instabilities when applied to more realistic systems and
for longer simulation times \citep{Berges2005,Berges2007,Aarts2010,Aarts2011,Gautier2012}.
Later a modification of the idea was proposed: to introduce complexified
configurations $x$ in path integrals and shift the integration contour
(domain) in order to make the oscillations of trajectory weights milder
\citep{Cristoforetti2012}. This is achieved by locating the stationary
points of complexified (analytically continued) action $S\left[x\right]$
and find the pathes (hypersurfaces) of steepest descent. These pathes
are called Lefschetz thimbles (LT). They have the property that the
phase (real part of $S\left[x\right]$) is constant on them. The idea
is to perform Monte Carlo sampling on Lefschetz thimbles \citep{Scorzato2015}.
The limitations of this approach follow from the fact that Lefschetz
thimble is a complicated curved hypersurface for which there is no
explicit relations. Nevertheless, there are promising results and
developments \citep{Cristoforetti2013,Renzo2015,Alexandru2016,Alexandru2016a,Alexandru2016b,Tsutsui2016}.
In a recently published work an idea was proposed to conduct Monte
Carlo simulations in the vicinity of Lefschetz thimble so that the
oscillations of trajectory weights are greatly supressed \citep{Alexandru2016c}. 

A common feature of complex Langevin and Lefschetz thimble methods
is that they start from a given discretized path integral on a given
integration domain and then just try to complexity it (to continuously
deform the integration domain into the complex plane). 

In this work we reconsider the standard derivation of path integrals
\citep{Zinn-Justin2005} and construct a discretized action where
the path integration domain can be deformed so that at least Gaussian
integrals become probabilistic. It turns out that there is rich freedom
when constructing discretized actions. Here we explore only one component
of this freedom, the integration contour (domain) independence, which
turns out to be richer than in Lefschetz thimble methods. We discuss
the relation of the analytical continuation approach with the Feynman's
arguments that it is imposible to simulate quantum dynamics on classical
local probabilistic computer \citep{Feynman1981}. Our approach is
based on work \citep{Rosales-Zarate2014}. As a particular application
of this general methodology, we consider a quantum system interacting
with Gaussian heat bath and show that probabilistic path integral
for the heat bath leads to a stochastic evolution of reduced system
state, analogous to what was derived from different considerations
in works \citep{Strunz1999,Stockburger2002,Shao2004,Lacroix2008,Hupin2010}. 

We believe that the rich freedom implicilty present in the discretized
path integral actions will help us to devise novel simulation algorithms
for real-time quantum dynamics, and to improve efficiency of the existing
complex Langevin and Lefschetz thimble methods.

\section{GENERALIZED PATH INTEGRALS}

In this section we demonstrate that the expression for discretized
path integral action is not unique, and we can exploit this fact in
order to make the Gaussian path integral real positive.

The propagation of bosonic quantum many-particle system is described
by evolution operator
\begin{equation}
\widehat{U}=e^{-i\left(t_{f}-t_{i}\right)\widehat{H}\left(\widehat{a}^{\dagger},\widehat{a}\right)}.\label{eq:unitary_evolution}
\end{equation}
The matrix element of $\widehat{U}$ between given initial $\Psi_{i}$
and final $\Psi_{f}$ wavefunctionals can be represented formally
as path integral

\begin{multline}
\left\langle f\right|\widehat{U}\left|i\right\rangle =\int D\left[\alpha,\alpha^{*}\right]\\
\times\exp\left\{ \int_{t_{i}}^{t_{f}}dt\int dx\left[\frac{1}{2}\left(\alpha\partial_{t}\alpha^{*}-\alpha^{*}\partial_{t}\alpha\right)-iH\left(\alpha,\alpha^{*}\right)\right]\right\} \\
\times\exp\left\{ \frac{1}{2}\int dx\left[\left|\alpha\left(x,t_{i}\right)\right|^{2}+\left|\alpha\left(x,t_{f}\right)\right|^{2}\right]\right\} \\
\times\Psi_{f}^{*}\left(\alpha\left(\cdot,t_{f}\right)\right)\Psi_{i}\left(\alpha\left(\cdot,t_{i}\right)\right).\label{eq:unitary_evolution_as_pi}
\end{multline}
Here, the second line can be interpreted as defining the canonical
structure of the evolution; the third line as convergence factor of
the integral; the last line specifies boundary conditions for the
matrix element. 

We recall that the standard derivation of path integral begins with
introduction of the holomorphic representation of quantum states.
Here, each quantum state with occupied modes $i_{1}\ldots i_{N}$
is mapped onto analytic function of variables $\alpha_{i_{1}}\ldots\alpha_{i_{N}}$
according to the rule
\begin{equation}
\Psi\left(\widehat{a}^{\dagger}\right)=\widehat{a}_{i_{1}}^{\dagger}\ldots\widehat{a}_{i_{N}}^{\dagger}\left|0\right\rangle \to\Psi\left(\alpha\right)=\alpha_{i_{1}}\ldots\alpha_{i_{N}},
\end{equation}
and superpositions of states are represented by superpositions of
analytic functions. It can be shown that in holomorphic picture the
inner product between any pair of quantum states is given by
\begin{multline}
\left\langle \Psi_{1}^{*}\left(\widehat{a}\right)\left|\Psi_{2}\left(\widehat{a}^{\dagger}\right)\right.\right\rangle =\int\prod_{k=1}^{M}\frac{d\alpha_{k}d\alpha_{k}^{*}}{2\pi i}\\
\times e^{-\sum_{k}\alpha_{k}\alpha_{k}^{*}}\left[\Psi_{1}\left(\alpha\right)\right]^{*}\Psi_{2}\left(\alpha\right).\label{eq:holomorphic_dot_product}
\end{multline}
The action of creation and annihilation operators is represented by
\begin{equation}
\widehat{a}_{i}^{\dagger}\Psi\left(\widehat{a}^{\dagger}\right)\to\alpha_{i}\Psi\left(\alpha\right),\,\,\,\widehat{a}_{i}\Psi\left(\widehat{a}^{\dagger}\right)\to\partial_{\alpha_{i}}\Psi\left(\alpha\right).
\end{equation}
Therefore, arbirary normally ordered operator $O\left(\widehat{a}^{\dagger},\widehat{a}\right)$
is represented as a differential operator $O\left(\alpha,\partial_{\alpha}\right)$.
However, in order to arrive at the path integral representation (\ref{eq:unitary_evolution_as_pi}),
we need to represent the action of $O\left(\widehat{a}^{\dagger},\widehat{a}\right)$
as integral kernel. We achieve this by employing integral automorphisms
of holomorphic representation: 
\begin{equation}
\Psi\left(\alpha\right)=\int\prod_{k=1}^{M}\frac{d\alpha_{k}^{\prime}d\alpha_{k}^{\prime*}}{2\pi i}\mathcal{A}\left(\alpha,\alpha^{\prime}\right)\Psi\left(\alpha^{\prime}\right).\label{eq:holomorphic_automorphism}
\end{equation}
The integral kernels $\mathcal{A}\left(\alpha,\alpha^{\prime}\right)$
contain considerable degree of freedom. Restricting ourselves to a
from which is Gaussian in $\alpha^{\prime}$, $\alpha^{\prime*}$,
we have the following family of $\mathcal{A}\left(\alpha,\alpha^{\prime}\right)$:
\begin{equation}
\mathcal{A}\left(\alpha,\alpha^{\prime}\right)=\det A\exp\left(-\sum_{kl}\left(\alpha_{k}^{\prime}-\alpha_{k}\right)A_{kl}\left(\alpha_{l}^{\prime*}-b_{l}\right)\right),
\end{equation}
where in general $A_{kl}$ and $b_{l}$ may depend on $\alpha$ and
$\alpha^{*}$. In the standard derivation of path integral one usually
sets $A_{kl}=\delta_{kl}$ and $b_{l}=0$ \citep{Zinn-Justin2005}.
The action of $\mathcal{A}$ is based on Wick theorem: $\alpha_{k}^{\prime}$
are replaced by their ``mean values'' $\alpha_{k}$; if $\alpha_{l}^{\prime*}$
were present then they would be replaced by $b_{l}$; if the pairs
$\alpha_{k}^{\prime*}\alpha_{l}^{\prime}$ were present then they
would be contracted by $\left(A^{-1}\right)_{kl}$. For the Wick theorem
to hold the integration countour (hypersurface) is irrelevant. The
only requirement is that all the Gaussian integrals coming from all
$\mathcal{A}$s should be convergent. The standard choice is that
the real $\alpha_{xk}^{\prime}$ and imaginary $\alpha_{yk}^{\prime}$
parts of $\alpha_{k}^{\prime}$ run along the real axis from $-\infty$
to $+\infty$. However we will assume that the contour is arbitrary
and $\alpha_{xk}^{\prime}$ and $\alpha_{yk}^{\prime}$ are independent
complex numbers. Therefore, instead of $\alpha_{k}^{*}$ we will employ
$\alpha_{k}^{\natural}=\alpha_{xk}-i\alpha_{yk}$. This way the normally
ordered operator $O\left(\widehat{a}^{\dagger},\widehat{a}\right)$
is represented as a kernel of integral operator
\begin{equation}
O\left(\widehat{a}^{\dagger},\widehat{a}\right)\to\mathcal{O}\left(\alpha,\alpha^{\prime}\right)=O\left(\alpha,\partial_{\alpha}\right)\mathcal{A}\left(\alpha,\alpha^{\prime}\right).
\end{equation}

Now we introduce the grid of time moments $t_{j}$, $j=0\ldots P$,
with time step $\Delta t$ and boundary conditions $t_{0}=t_{i}$
and $t_{P}=t_{f}$. The holomorphic representations of initial and
final quantum states are 
\begin{equation}
\left|\Psi_{i}\right\rangle \to\Psi_{0}\left(\alpha_{0}\right),\,\,\,\left|\Psi_{f}\right\rangle \to\Psi_{P}\left(\alpha_{P}\right).
\end{equation}
The propagation during time interval $\left[t_{j},t_{j+1}\right]$,
\begin{equation}
\left|\Psi_{j+1}\right\rangle =\left(1-i\Delta t\widehat{H}\left(\widehat{a}^{\dagger},\widehat{a}\right)+O\left(\Delta t^{2}\right)\right)\left|\Psi_{j}\right\rangle ,
\end{equation}
in holomorphic picture assumes the form
\begin{multline}
\Psi_{j+1}\left(\alpha\left(j+1\right)\right)=\int\prod_{k}\frac{d\alpha_{xk}\left(j\right)d\alpha_{yk}\left(j\right)}{2\pi i}\\
\times\exp\left[-\sum_{k}\alpha_{k}^{\natural}\left(j\right)\left(\alpha_{k}\left(j+1\right)-\alpha_{k}\left(j\right)\right)\right.\\
\left.-i\Delta tH\left(\alpha\left(j+1\right),\alpha^{\natural}\left(j\right)\right)+O\left(\Delta t^{2}\right)\vphantom{\sum_{k}\alpha_{k}^{\natural}\left(j\right)\left(\alpha_{k}\left(j+1\right)-\alpha_{k}\left(j\right)\right)}\right]\Psi_{j}\left(\alpha\left(j\right)\right).
\end{multline}
Here we have employed the standard form of automorphism \citep{Zinn-Justin2005}.
Using this recurrent relation between the neighboring $\Psi_{j}\left(\alpha\left(j\right)\right)$
and the inner product formula (\ref{eq:holomorphic_dot_product}),
we construct the discretized path integral. For a general Gaussian
problem we have
\begin{multline}
\left\langle f\right|\widehat{U}\left|i\right\rangle =\int\prod_{k,j}\frac{d\alpha_{xk}\left(j\right)d\alpha_{yk}\left(j\right)}{2\pi i}\\
\times\exp\left[-\sum_{kjlp}\alpha_{k}\left(j\right)\left(G^{-1}\right)_{kj,lp}\alpha_{l}^{\natural}\left(p\right)\right].\label{eq:gaussian_pi}
\end{multline}
We can deform the integration hypersurface for $\alpha_{xk}$ and
$\alpha_{yk}$ such that the integrand becomes real positive. Indeed,
suppose we have some factorization of $G$,
\begin{equation}
G=UV^{*}.\label{eq:factorization_of_green}
\end{equation}
Then we can choose such integration hypersurface that the following
variables become complex conjugate:
\begin{equation}
\gamma_{k}=\sum_{lj}\alpha_{l}\left(j\right)\left(V^{-1}\right)_{lj,k}^{*},\,\,\,\gamma_{k}^{*}=\sum_{lj}\left(U^{-1}\right)_{k,lj}\alpha_{l}^{\natural}\left(j\right).\label{eq:conjugate_variables}
\end{equation}
This way we obtain the following probabilistic path integral:
\begin{multline}
\left\langle f\right|\widehat{U}\left|i\right\rangle =\int\prod_{kj}\frac{d\alpha_{xk}\left(j\right)d\alpha_{yk}\left(j\right)}{2\pi i}\\
\times\exp\left[-\sum_{k}\gamma_{k}\left(\alpha_{x},\alpha_{y}\right)\gamma_{k}^{*}\left(\alpha_{x},\alpha_{y}\right)\right].\label{eq:probabilistic_path_integral}
\end{multline}

The main result of this section is that such properties of holomorphic
representation as contour-independence and the existence of a rich
set of automorphisms allows one to solve the sign problem at least
for Gaussian systems. The case of interacting systems (e.g. the presence
of quartic terms in the action) deserves further study. 

If we change the variables from $\alpha,\alpha^{\natural}$ to $\gamma,\gamma^{*}$,
we consider them as new independent classical random quantities with
the following statisitcs:
\begin{equation}
\overline{\gamma_{k}^{*}\gamma_{l}}=\delta_{kl}.
\end{equation}
From Eq. (\ref{eq:conjugate_variables}) we have:
\begin{equation}
\alpha=V^{\dagger}\gamma,\,\,\,\alpha^{\natural}=U\gamma^{*}.\label{eq:change_to_stochastic_variables}
\end{equation}
Therefore, the covariance of $\alpha,\alpha^{\natural}$ is
\begin{equation}
\overline{\alpha^{\natural}\otimes\alpha^{T}}=U\overline{\gamma^{*}\otimes\gamma^{T}}V^{*}=UV^{*}=G
\end{equation}
as it should be for the original path integral (\ref{eq:gaussian_pi}). 

\section{MONTE CARLO SIMULATION OF BELL STATES}

In the preceeding section we have shown that it is possible to represent
arbitrary quantum Gaussian evolution as a classical stochastic process.
This may seem somewhat surprising, especially in light of classical
work of Feynman \citep{Feynman1981} where he argues that it is impossible
to efficiently simulate quantum evolution on a classical local probabilistic
computer. Since Feynman in his work considers the two-photon correlation
experiment, here we provide computations for this case. 

Let us consider evolution under the parametric down-conversion (PDC)
process, which has the following effective Hamiltonian \citep{Reid1986}
\begin{equation}
\widehat{H}=i\kappa\left(\widehat{a}_{1}^{\dagger}\widehat{b}_{1}^{\dagger}-\widehat{a}_{1}\widehat{b}_{1}\right)+i\kappa\left(\widehat{a}_{2}^{\dagger}\widehat{b}_{2}^{\dagger}-\widehat{a}_{2}\widehat{b}_{2}\right),\label{eq:PDC_hamiltonian}
\end{equation}
where $1$ and $2$ denote the two orthogonal polarizations; the photonic
modes $a_{i}$ propagate ``to the left'' wing of the experiment,
and the photonic modes $b_{i}$ propagate ``to the right'' wing
of the experiment. At each wing of the experiment there are polarizers
$\mathcal{A}$ and $\mathcal{B}$ at the angles $\theta$ and $\phi$
correspondingly. We measure the transmitted $\widehat{c}_{+}$ and
reflected $\widehat{c}_{-}$ modes at polarizer $\mathcal{A}$ 
\begin{equation}
\widehat{c}_{+}=\widehat{a}_{1}\cos\theta+\widehat{a}_{2}\sin\theta,
\end{equation}
\begin{equation}
\widehat{c}_{-}=-\widehat{a}_{1}\sin\theta+\widehat{a}_{2}\cos\theta,
\end{equation}
and the transmitted $\widehat{d}_{+}$ and reflected $\widehat{d}_{-}$
modes at polarizer $\mathcal{B}$,
\begin{equation}
\widehat{d}_{+}=\widehat{b}_{1}\cos\phi+\widehat{b}_{2}\sin\phi,
\end{equation}
\begin{equation}
\widehat{d}_{-}=-\widehat{b}_{1}\sin\phi+\widehat{b}_{2}\cos\phi.
\end{equation}
We use the Clauser-Horne-Bell inequality with intensity moments \citep{Reid1986}:
\begin{multline}
S_{\textrm{CH}}\\
=\frac{I_{++}^{AB}\left(\theta,\phi\right)-I_{++}^{AB}\left(\theta,\phi^{\prime}\right)+I_{++}^{AB}\left(\theta^{\prime},\phi\right)+I_{++}^{AB}\left(\theta^{\prime},\phi^{\prime}\right)}{I_{+}^{A}\left(\theta^{\prime}\right)+I_{+}^{B}\left(\phi\right)}\\
\leq1.\label{eq:CHB_inequality}
\end{multline}
Here
\begin{equation}
I_{+}^{A}\left(\theta^{\prime}\right)=\left\langle \widehat{c}_{+}^{\dagger}\widehat{c}_{+}\left(\widehat{d}_{+}^{\dagger}\widehat{d}_{+}+\widehat{d}_{-}^{\dagger}\widehat{d}_{-}\right)\right\rangle ,\label{eq:p_a}
\end{equation}
is the correlation of the intensity of photons which are transmitted
through the polarizer $\mathcal{A}$ at the angle $\theta^{\prime}$
with the full intensity of photons which are incident on the polarizer
$\mathcal{B}$; 
\begin{equation}
I_{+}^{B}\left(\phi\right)=\left\langle \widehat{d}_{+}^{\dagger}\widehat{d}_{+}\left(\widehat{c}_{+}^{\dagger}\widehat{c}_{+}+\widehat{c}_{-}^{\dagger}\widehat{c}_{-}\right)\right\rangle \label{eq:p_b}
\end{equation}
is the symmetrically mirrored quantity;
\begin{equation}
I_{++}^{AB}\left(\theta,\phi\right)=\left\langle \widehat{c}_{+}^{\dagger}\widehat{c}_{+}\widehat{d}_{+}^{\dagger}\widehat{d}_{+}\right\rangle \label{eq:p_ab}
\end{equation}
is the correlation between the intensities of the photons which are
transmitted through the polarizers $\mathcal{A}$ and $\mathcal{B}$
simultaneously, at the angles $\theta$ and $\phi$ correspondingly.
The maximal violation of (\ref{eq:CHB_inequality}) should occur at
$\theta=0^{\circ}$, $\theta^{\prime}=45^{\circ}$, $\phi=22.5^{\circ}$,
$\phi^{\prime}=67.5^{\circ}$ when $S_{\textrm{CH}}\approx1.2$. The
Keldysh Green functions $G$ for Hamiltonian Eq. (\ref{eq:PDC_hamiltonian})
can be computed analytically by switching to the normal modes of this
Hamiltonian. Factorizing $G$ according to (\ref{eq:factorization_of_green})
and introducing the classical stochastic processes ($c_{\pm}$, $c_{\pm}^{\natural}$,
$d_{\pm}$, $d_{\pm}^{\natural}$) according to (\ref{eq:change_to_stochastic_variables}),
we compute the intensity moments Eqs. (\ref{eq:p_a})-(\ref{eq:p_ab})
as classical expectations $\textrm{E}_{\beta,\beta*}\left[c_{+}\left(t\right)c_{+}^{\natural}\left(t\right)\left(d_{+}\left(t\right)d_{+}^{\natural}\left(t\right)+d_{-}\left(t\right)d_{-}^{\natural}\left(t\right)\right)\right]$,
$\textrm{E}_{\beta,\beta*}\left[d_{+}\left(t\right)d_{+}^{\natural}\left(t\right)\left(c_{+}\left(t\right)c_{+}^{\natural}\left(t\right)+c_{-}\left(t\right)c_{-}^{\natural}\left(t\right)\right)\right]$
and $\textrm{E}_{\beta,\beta*}\left[c_{+}\left(t\right)c_{+}^{\natural}\left(t\right)d_{+}\left(t\right)d_{+}^{\natural}\left(t\right)\right]$
correspondingly.

\begin{figure}
\includegraphics[scale=0.4]{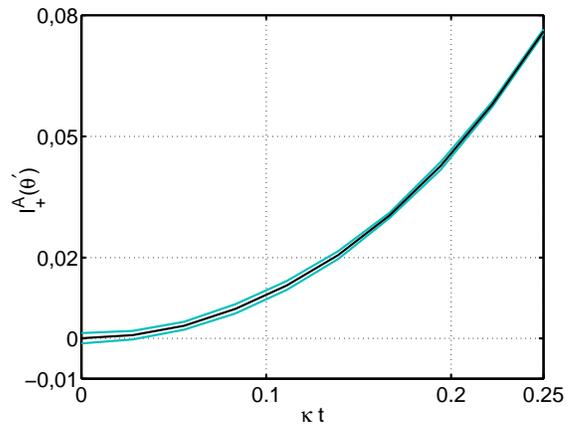}

\caption{\label{fig:I_A_PRIME}Correlation $I_{+}^{A}\left(\theta^{\prime}\right)$
of the intensity of transmitted photons through polarizer $\mathcal{A}$
with the full intensity of photons through plolarizer $\mathcal{B}$.
The central black line is the calculation result. The lower and the
upper color lines denote the standard deviation of the result. }

\end{figure}
\begin{figure}
\includegraphics[scale=0.4]{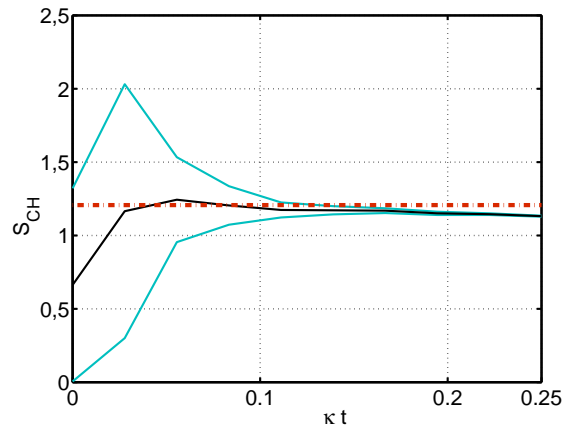}

\caption{\label{fig:S_CH}Left-hand side $S_{\textrm{CH}}$ of the Clauser-Horne-Bell
inequality Eq. (\ref{eq:CHB_inequality}) for evolution under the
PDC Hamiltonian Eq. (\ref{eq:PDC_hamiltonian}) starting from vacuum
state. The dash-dotted red line is the maximal violation $S_{\textrm{CH}}\approx1.2$.
The meaning of the other lines is the same as in Fig. \ref{fig:I_A_PRIME}.}
\end{figure}
In order to give an idea of the calculated intensity correlations,
on Fig. \ref{fig:I_A_PRIME} we present the results for $I_{+}^{A}\left(\theta^{\prime}\right)$.
The dynamical evolution of the left hand side of the Clauser-Horne-Bell
inequality Eq. (\ref{eq:CHB_inequality}) is presented on Fig. \ref{fig:S_CH}.
The large variance of computed $S_{\textrm{CH}}$ near $\kappa t=0$
is because all the intensity correlations become exponentially small
when $\kappa t\to0$ (see e.g. Fig. \ref{fig:I_A_PRIME}). In fact,
this large variance reflects the fact that such an experimental setup
becomes inefficient for the Bell-type tests when $\kappa t\ll1$,
since the detection of photons becomes exponetially-seldom event.
The computation is carried at the angles of maximal violation of $S_{\textrm{CH}}$
(dash-dotted line on Fig. \ref{fig:S_CH}). Neverthless, the value
of maximal violation is not reached and $S_{\textrm{CH}}$ is decreased
with time because during the evolution under the PDC Hamiltonian (\ref{eq:PDC_hamiltonian})
the increasing multiphoton contribution to the Bell state makes it
non-ideal. 

\section{OPEN SYSTEMS AND PROBABILISTIC PATH INTEGRAL FOR THE BATH}

The Monte Carlo computation of polynomial averages for a Gaussian
system with known Green function represents purely methodological
interest since in this case we can apply Wick theorem. However the
problem of the dynamics of an open quantum system interacting with
a Gaussian bath is already nontrivial. Such a problem emerges in different
fields of research, ranging from open systems and decoherence theory
\citep{Leggett1987} to computation methods for correlated systems
(Anderson impurity model and the dynamical mean field theory \citep{Georges1996}).
In this section we discuss how the probabilistic path integral leads
to computational recipes for this kind of problems. This problem was
already treated from a different point of view in works \citep{Strunz1999,Stockburger2002,Shao2004,Lacroix2008,Hupin2010},
however our consideration provides us with a different methodological
perspective.

Suppose we have a system 
\begin{equation}
\widehat{H}=\widehat{H}_{\textrm{q}}+h\left(\widehat{b}^{\dagger}\widehat{a}+\widehat{b}\widehat{a}^{\dagger}\right)+\widehat{H}_{\textrm{b}},
\end{equation}
where $\widehat{H}_{\textrm{b}}$ is a Gaussian bath with known contour
Green function 
\begin{equation}
G_{\textrm{b}}\left(\tau-\tau^{\prime}\right)=\left\langle 0_{\textrm{b}}\left|\mathcal{T}\widehat{a}\left(\tau\right)\widehat{a}^{\dagger}\left(\tau^{\prime}\right)\right|0_{\textrm{b}}\right\rangle \label{eq:bath_Green}
\end{equation}
for its coupled degree of freedom $\widehat{a}$. We assume that the
initial state of the system and of the bath is factorized as $\left|\Psi_{\textrm{ini}}\right\rangle \left|0_{\textrm{b}}\right\rangle $,
where $\left|0_{\textrm{b}}\right\rangle $ is the vacuum of the bath.
We also suppose that we are dealing with a certain dynamical problem
up to a real time $T$ for which appropriate contour (e.g. Keldysh)
was chosen with forward $\mathcal{C}_{+}$ and backward $\mathcal{C}_{-}$
branches. The system $\widehat{H}_{\textrm{q}}$ may be arbitrary
yet numericaly tractable (that is whose Hilbert space dimension is
not very large). 

Let us describe the evolution of the reduced state of the system.
Since in a reduced description we average over the bath, we represent
it as a path intergal. This way each occurence of the operators $\widehat{a}$
and $\widehat{a}^{\dagger}$ is replaced by $c$-number stochastic
fields $\alpha^{*}\left(\tau\right)$ and $\alpha\left(\tau\right)$.
Now, we deform the integration domain to make the bath path integral
probabilistic. The price paid is that instead of complex conjugated
fields $\alpha^{*}\left(\tau\right)$ and $\alpha\left(\tau\right)$
we now deal with non-conjugated fields $\alpha^{\natural}\left(\tau\right)$
and $\alpha\left(\tau\right)$, and the corresponding evolution of
the system is not unitary. The obtained Hamiltonian in the interaction
picture with respect to the bath has the following stochastic form
\begin{equation}
\widehat{H}_{\textrm{stoch}}\left(\alpha_{\pm}\left(\tau\right),\alpha_{\pm}^{\natural}\left(\tau\right)\right)=\widehat{H}_{\textrm{q}}+h\left(\alpha_{\pm}\left(\tau\right)\widehat{b}+\alpha_{\pm}^{\natural}\left(\tau\right)\widehat{b}^{\dagger}\right),\label{eq:stochastic_hamiltonian}
\end{equation}
which governs the evolution of system's states $\left|\Psi_{+}\left(\tau\right)\right\rangle $
on the forward branch and $\left\langle \Psi_{-}\left(\tau\right)\right|$
on the backward branch according to the following stochastic differential
equations:
\begin{multline}
d\left|\Psi_{+}\left(\tau\right)\right\rangle \\
=-idt\left[\widehat{H}_{\textrm{q}}+h\left(\alpha_{+}\left(\tau\right)\widehat{b}+\alpha_{+}^{\natural}\left(\tau\right)\widehat{b}^{\dagger}\right)\right]\\
\times\left|\Psi_{+}\left(\tau\right)\right\rangle ,\label{eq:sde_forward}
\end{multline}
\begin{multline}
d\left\langle \Psi_{-}\left(\tau\right)\right|=\left\langle \Psi_{-}\left(\tau\right)\right|\\
\times idt\left[\widehat{H}_{\textrm{q}}+h\left(\alpha_{-}\left(\tau\right)\widehat{b}+\alpha_{-}^{\natural}\left(\tau\right)\widehat{b}^{\dagger}\right)\right].\label{eq:sde_backward}
\end{multline}
The average value of system's observable is computed as
\begin{equation}
\left\langle \widehat{O}\left(\tau\right)\right\rangle =\overline{\left\langle \Psi_{-}\left(\tau\right)\right|\widehat{O}\left|\Psi_{+}\left(\tau\right)\right\rangle },\label{eq:average_of_observable}
\end{equation}
 If the system is initially in the state $\left|\Psi_{\textrm{in}}\right\rangle $,
the equations (\ref{eq:sde_forward}) and (\ref{eq:sde_backward})
have the initial conditions $\Psi_{\pm}\left(0\right)=\Psi_{\textrm{in}}$. 

Now let us discuss how to generate the field configurations $\alpha^{\natural}\left(\tau\right)$
and $\alpha\left(\tau\right)$. The most straightforward way would
be to apply the relations Eq. (\ref{eq:change_to_stochastic_variables})
and to sample $\gamma_{l}$ from the standard compex normal distribution.
The singular value decomposition of Green function
\begin{equation}
G=U_{\textrm{s}}\Sigma V_{\textrm{s}}^{*}
\end{equation}
provides us with $U=U_{\textrm{s}}\sqrt{\Sigma}$ and $V^{*}=\sqrt{\Sigma}V_{\textrm{s}}^{*}$.
As preliminary calculations show, this approach is exact and converging
when the simulation sample size is increased. Neverthless, the statistical
variance of Monte Carlo results grows exponentially with time. The
reason of this behaviour becomes clear if we look at the time dependence
of system's state norm (overlap). Since the norm is conserved and
the simulation is exact, we have
\begin{equation}
\overline{\left\langle \Psi_{-}\left(\tau\right)\left|\Psi_{+}\left(\tau\right)\right.\right\rangle }=1.
\end{equation}
Nevertheless, for a given realization of field trajectories $\alpha^{\natural}\left(\tau\right)$
and $\alpha\left(\tau\right)$ we have 
\begin{multline}
\left\langle \Psi_{-}\left(\tau\right)\left|\Psi_{+}\left(\tau\right)\right.\right\rangle =\left\langle \Psi_{-}\left(0\right)\left|\Psi_{+}\left(0\right)\right.\right\rangle \\
\times\exp\left\{ i\intop_{0}^{\tau}d\tau^{\prime}\overline{b}\left(\tau^{\prime}\right)\left[\alpha_{-}\left(\tau^{\prime}\right)-\alpha_{+}\left(\tau^{\prime}\right)\right]\right.\\
\left.+i\intop_{0}^{\tau}d\tau^{\prime}\overline{b}^{+}\left(\tau^{\prime}\right)\left[\alpha_{-}^{\natural}\left(\tau^{\prime}\right)-\alpha_{+}^{\natural}\left(\tau^{\prime}\right)\right]\right\} ,\label{eq:overlap_expression-1}
\end{multline}
where $\overline{b}\left(\tau\right)$ and $\overline{b}^{+}\left(\tau\right)$
are the ``instantaneous'' mean values of system annihilation and
creation operators:
\begin{equation}
\overline{b}\left(\tau\right)=\frac{\left\langle \Psi_{-}\left(\tau\right)\right|\widehat{b}\left|\Psi_{+}\left(\tau\right)\right\rangle }{\left\langle \Psi_{-}\left(\tau\right)\left|\Psi_{+}\left(\tau\right)\right.\right\rangle },\label{eq:mean_s}
\end{equation}
\begin{equation}
\overline{b}^{+}\left(\tau\right)=\frac{\left\langle \Psi_{-}\left(\tau\right)\right|\widehat{b}^{\dagger}\left|\Psi_{+}\left(\tau\right)\right\rangle }{\left\langle \Psi_{-}\left(\tau\right)\left|\Psi_{+}\left(\tau\right)\right.\right\rangle }.\label{eq:mean_s_dagger}
\end{equation}
From Eq. (\ref{eq:overlap_expression-1}) we see that the normalization
of system's state fluctuates exponentially with time and this is the
reason for the growth of the variance of Monte Carlo results. At least
for harmonic system $\widehat{H}_{\textrm{q}}=\varepsilon\widehat{b}^{\dagger}\widehat{b}$
it is possible to overcome this problem in the following way. Firstly,
we employ the trick analogous to \citep{Stockburger2002}: we change
the variables in the bath path integral according to
\begin{equation}
\alpha\left(\tau\right)\to\alpha\left(\tau\right)+f\left(\tau\right),\label{eq:1_transformation}
\end{equation}
\begin{equation}
\alpha^{\natural}\left(\tau\right)\to\alpha^{\natural}\left(\tau\right)+f^{\natural}\left(\tau\right),\label{eq:2_transformation}
\end{equation}
where $f\left(\tau\right)$ and $f^{\natural}\left(\tau\right)$ are
retarded functionals of the fields $\alpha^{\natural}\left(\tau\right)$
and $\alpha\left(\tau\right)$:
\begin{equation}
f\left(\tau\right)=-i\intop_{\mathcal{C}}d\tau^{\prime}\overline{b}\left(\tau^{\prime}\right)G\left(\tau,\tau^{\prime}\right),\label{eq:noise_addition}
\end{equation}
\begin{equation}
f^{\natural}\left(\tau\right)=-i\intop_{\mathcal{C}}d\tau^{\prime}\overline{b}^{+}\left(\tau^{\prime}\right)G\left(\tau^{\prime},\tau\right).\label{eq:conjugated_noise_addition}
\end{equation}
Here according to Eqs. (\ref{eq:mean_s}) and (\ref{eq:mean_s_dagger}),
$\overline{b}^{+}\left(\tau^{\prime}\right)$ and $\overline{b}\left(\tau^{\prime}\right)$
take the same values on $\mathcal{C}_{+}$ and $\mathcal{C}_{-}$.
The functional Jacobian matrices of the transformation (\ref{eq:1_transformation})-(\ref{eq:2_transformation})
are lower triangular with unit diagonal, thus their determinant is
unity, and the integration measure is not changed. Then it can be
shown that the overlap factor Eq. (\ref{eq:overlap_expression-1})
is compensated. The modified stochastic fields Eqs. (\ref{eq:1_transformation})-(\ref{eq:2_transformation})
are sampled as
\begin{equation}
\alpha\left(\tau_{l}\right)=\sum_{k}V_{lk}^{\dagger}\gamma_{k}+f\left(\tau_{l}\right),\label{eq:modified_noise}
\end{equation}
\begin{equation}
\alpha^{\natural}\left(\tau_{l}\right)=\sum_{k}U_{lk}\gamma_{k}^{*}+f^{\natural}\left(\tau_{l}\right),\label{eq:conjugated_modified_noise}
\end{equation}
where as previously $\gamma_{k}$ are sampled from the standard normal
distribution. Now in principle we could solve the stochastic equations
(\ref{eq:sde_forward})-(\ref{eq:sde_backward}) with this modified
noise and evaluate the observables as 
\begin{equation}
\left\langle \widehat{O}\left(\tau\right)\right\rangle =\overline{\left[\frac{\left\langle \Psi_{-}\left(\tau\right)\right|\widehat{O}\left|\Psi_{+}\left(\tau\right)\right\rangle }{\left\langle \Psi_{-}\left(\tau\right)\left|\Psi_{+}\left(\tau\right)\right.\right\rangle }\right]},
\end{equation}
where the denominator accounts for the fact that we have compensated
for the norm. Let us assume that the initial system state $\Psi_{\textrm{ini}}$
is a coherent one. Under Gaussian evolution the coherent state conserves
its form. Therefore, at any time moment $\tau$ we have
\begin{equation}
\left|\Omega_{\pm}\left(\tau\right),b_{\pm}\left(\tau\right)\right\rangle =\exp\left[\Omega_{\pm}\left(\tau\right)+b_{\pm}\left(\tau\right)\widehat{b}^{\dagger}\right]\left|0\right\rangle ,
\end{equation}
where $\Omega_{\pm}\left(\tau\right)$ is a normalization and $b_{\pm}\left(\tau\right)$
is a displacement. In terms of $\Omega_{\pm}\left(\tau\right)$ and
$b_{\pm}\left(\tau\right)$, the stochastic equations Eqs. (\ref{eq:sde_forward})-(\ref{eq:sde_backward})
take the following form:
\begin{equation}
db_{+}\left(\tau\right)=-i\varepsilon b_{+}\left(\tau\right)d\tau-ih\alpha_{+}^{\natural}\left(\tau\right)d\tau,\label{eq:s_equation}
\end{equation}
\begin{equation}
d\Omega_{+}\left(\tau\right)=-ihb_{+}\left(\tau\right)\alpha_{+}\left(\tau\right)d\tau,\label{eq:omega_equation}
\end{equation}
and also the adjoint equations for $\Omega_{-}\left(\tau\right)$,
$b_{-}\left(\tau\right)$. The stochastic term in equations for $\Omega_{\pm}$
{[}Eq. (\ref{eq:omega_equation}){]} is multiplicative and this fact
could lead to exponential growth of variance. However since we divide
by overlap $\left\langle \Omega_{-},b_{-}\left(\tau\right)\left|\Omega_{+},b_{+}\left(\tau\right)\right.\right\rangle $,
the dependence on $\Omega_{\pm}\left(\tau\right)$ is compensated,
and we can drop the equation (\ref{eq:omega_equation}). The ``instantaneous''
mean values of system annihilation and creation operators Eqs. (\ref{eq:mean_s})
- (\ref{eq:mean_s_dagger}) assume the following simple form: 
\begin{equation}
\overline{b}\left(\tau\right)=b_{+}\left(\tau\right),\,\,\,\overline{b}^{+}\left(\tau\right)=b_{-}^{*}\left(\tau\right).\label{eq:mean_for_coherent_states}
\end{equation}
The resulting simulation procedure is defined by equation of motion
Eq. (\ref{eq:s_equation}), adjoint equation for $b_{-}\left(\tau\right)$,
the noise is constructed according to Eq. (\ref{eq:mean_for_coherent_states}),
(\ref{eq:noise_addition})-(\ref{eq:conjugated_modified_noise}).
Since any state can be represented as a mixture of coherent states,
initial conditions of a more general form can be treated by the same
procedure.

\begin{figure}
\includegraphics[scale=0.4]{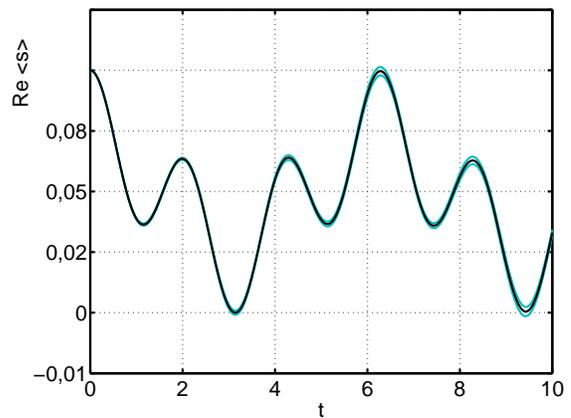}

\caption{\label{fig:real_s}Time evolution of evaluated $\textrm{Re}\left\langle \widehat{b}\right\rangle $
starting from coherent state with $b_{\pm}\left(0\right)=1$. The
meaning of the lines is the same as in Fig. \ref{fig:I_A_PRIME}. }

\end{figure}

In Fig. \ref{fig:real_s} we present the resutls of calculations for
a harmonic system with $\varepsilon=1$ which is coupled to one harmonic
mode $\widehat{H}_{\textrm{b}}=\omega\widehat{a}^{\dagger}\widehat{a}$
with $\omega=1$ and $h=2$. It is seen that the stochastic computation
correctly reproduces the beat due to coupling to the harmonic bath
mode. 

This recipe works only for a harmonic system which is in a Gaussian
state. If the system were not harmonic, it would be difficult to introduce
coherent states, and it would not be possible to compensate the multiplicative
noise terms like those in Eq. (\ref{eq:omega_equation}). If the state
was not Gaussian (coherent), then we had to use general formulas Eq.
(\ref{eq:mean_s})-(\ref{eq:mean_s_dagger}) which are numerically
unstable. The case of non-Gaussian evolution deserves further study.

\section{DISCUSSION\label{sec:CONCLUSION}}

In his classical paper \citep{Feynman1981} Feynman argues that the
efficient classical simulation of quantum dynamics is possible only
on local causal probabilistic computer. The possibility of such simulation
would imply the possibility of ``hidden parameters'' interpretation
of quantum mechanics, thus it is impossible. Nevertheless, as we have
discussed in this work, we can escape this limitation by noting that
Bell's theorem prohibits only ``hidden parameters'' interpretation
in terms of physical probabilies and physical states. If we make the
trajectories of states unphysical by analytic continuation of field
variables then at least for Gaussian systems we obtain probabilistic
path integrals. The latter can be sampled by Monte Carlo methods.
In different terms and context, similar conclusions have been drawn
in \citep{Rosales-Zarate2014}.

The situation when anharmonic interaction terms are present (e.g.
quartic in creation/annihilation operators) deserves further study.
In principle is is known that the deformation of contour which eliminates
the oscillation of complex phase is possible: this is the main result
of the Lefschetz thimbles approach \citep{Scorzato2015}. The open
question is to find such a deformed contour that the Monte Carlo sampling
along it is not prohibitively computationally complex. It is likely
that there is no universal answer to this question for a general nonlinear
quantum dynamical problem. Nevertheless, we believe that the freedom
in the derivation of discretized path integral action may be usefull
to increase the efficiency of simulation algorithms for specific classes
of dynamical problems.
\begin{acknowledgments}
The study was funded by the RSF, grant 16-42-01057. 
\end{acknowledgments}

\end{document}